\documentclass[12pt]{article}
\usepackage{graphicx,amsmath,amssymb}

\newlength{\dinwidth}
\newlength{\dinmargin}
\renewcommand{\vec}[1]{\boldsymbol{#1}}
\newcommand{\dif}{\mathrm{d}}
\newcommand{\diff}[1]{\frac{\mathrm{d}#1}{#1}}
\newcommand{\xB}{x_{\scriptscriptstyle{B}}}

\begin{document}
\titlepage
\begin{flushright}
  IPPP/03/50     \\
  DCPT/03/100    \\
  9th November 2004   \\
\end{flushright}

\vspace*{0.5cm}

\begin{center}

  {\Large \bf Unintegrated parton distributions and\\[1ex] electroweak boson production at hadron colliders}

  \vspace*{1cm}

  \textsc{G. Watt$^a$, A.D. Martin$^a$ and M.G. Ryskin$^{a,b}$} \\

  \vspace*{0.5cm}

  $^a$ Institute for Particle Physics Phenomenology, University of Durham, DH1 3LE, UK \\
  $^b$ Petersburg Nuclear Physics Institute, Gatchina, St.~Petersburg, 188300, Russia

\end{center}

\vspace*{0.5cm}

\begin{abstract}
  We describe the use of doubly-unintegrated parton distributions in hadron-hadron collisions, using the $(z,k_t)$-factorisation prescription where the transverse momentum of the incoming parton is generated in the last evolution step.  We apply this formalism to calculate the transverse momentum ($P_T$) distributions of produced $W$ and $Z$ bosons and compare the predictions to Tevatron Run 1 data.  We find that the observed $P_T$ distributions can be generated almost entirely by the leading order $q_1\,q_2\to W,Z$ subprocesses, using known and universal doubly-unintegrated quark distributions.  We also calculate the $P_T$ distribution of the Standard Model Higgs boson at the LHC, where the dominant production mechanism is by gluon-gluon fusion.
\end{abstract}

\section{Introduction}

At present, it is not straightforward to describe the transverse momentum ($P_T$) distributions of electroweak bosons produced in hadron-hadron collisions.  In the usual collinear approximation, the transverse momentum of the incoming partons is neglected and so, for the Born level subprocesses $q_1\,q_2\to V$ (where $V = \gamma^*,W,Z$) or $g_1\,g_2\to H$, the transverse momentum of the final electroweak boson is zero.  Therefore, initial-state QCD radiation is necessary to generate the $P_T$ distributions.  Both the leading order (LO) and next-to-leading order (NLO) differential cross sections diverge for $P_T\ll M_{V,H}$, with terms proportional to $\ln (M_{V,H}/P_T)$ appearing due to soft and collinear gluon emission, requiring resummation to achieve a finite $P_T$ distribution.

Traditional calculations combine fixed-order perturbation theory at high $P_T$ with either analytic resummation or numerical parton shower formalisms at low $P_T$, with some matching criterion to decide when to switch between the two.  In addition, a parameterisation is needed to account for non-perturbative effects at the lowest $P_T$ values.  Analytic resummation can be performed either in the transverse momentum space (see, for example, \cite{Ellis:1997ii}) or in the Fourier conjugate impact parameter space (see, for example, \cite{Balazs:1997xd}).

An alternate description is provided in terms of \emph{unintegrated} parton distribution functions (UPDFs), where each incoming parton carries its own transverse momentum $k_t$, so that the subprocesses $q_1\,q_2\to V$ and $g_1\,g_2\to H$ already generate the LO $P_T$ distributions in the so-called $k_t$-factorisation approach.\footnote{For an introduction to $k_t$-factorisation, see \cite{smallx}.}  It has been shown in \cite{Kwiecinski} that UPDFs obtained from an approximate solution of the CCFM evolution equation \cite{CCFM} embody the conventional soft gluon resummation formulae.

The UPDFs that we use are obtained from the familiar DGLAP-evolved PDFs determined from a global parton analysis of deep-inelastic and related hard-scattering data.  The transverse momentum of the parton is generated entirely in the \emph{last} evolution step \cite{Kimber:2000,Kimber:2001sc}.  Angular-ordering constraints are imposed which regulate the singularities arising from soft gluon emission, while the virtual terms in the DGLAP equation are resummed into Sudakov form factors.\footnote{It was found in \cite{Kimber:2001sc} that the `last-step' prescription gave similar results whether the input PDFs were evolved with the DGLAP equation or with a unified BFKL-DGLAP equation \cite{Kwiecinski:1997ee}, indicating that angular ordering is more important than small-$x$ effects.}  In \cite{Watt} it was shown that it is necessary to extend the `last-step' formalism of \cite{Kimber:2001sc} to consider \emph{doubly}-unintegrated parton distribution functions (DUPDFs) in order to preserve the exact kinematics.  It was demonstrated that the main features of conventional higher order calculations can be accounted for within a much simpler theoretical framework, named $(z,k_t)$-factorisation.\footnote{Here, $z$ is the splitting fraction associated with the last evolution step, where a parton with (light-cone) momentum fraction $x/z$ splits to a final parton with (light-cone) momentum fraction $x$.  The idea is an extension of the original DDT formula \cite{Dokshitzer:hw}; however, in comparison with \cite{Dokshitzer:hw} we go beyond the double leading logarithmic approximation (DLLA) and account for the precise kinematics of the two incoming partons, as well as the angular ordering of emitted gluons.}

Strictly speaking, the integrated PDFs used as input to the last evolution step should themselves be determined from a new global fit to data using the $(z,k_t)$-factorisation approach.  For the present paper, we take the input PDFs from a global fit to data using the conventional collinear approximation \cite{Martin:2002dr}.  This treatment is adequate for these initial investigations.  However, we expect it to lower our predictions for quark-initiated processes by $\sim$10\% compared to the case where the input PDFs are determined from a global fit using the $(z,k_t)$-factorisation approach.  We will illustrate this point in Section 5 by comparing predictions for the proton structure function $F_2$ in the collinear approximation and in the $(z,k_t)$-factorisation formalism.

The `last-step' prescription has some features in common with the initial-state parton shower algorithms implemented in Monte Carlo event generators (for a recent review, see \cite{Dobbs:2004qw}) such as the DGLAP-based \textsc{herwig} \cite{Herwig} and \textsc{pythia} \cite{Pythia} programs and the CCFM-based \textsc{cascade} \cite{Cascade} program.  The main advantage of our approach is that we use simple analytic formulae which implement the crucial physics in a transparent way, without the additional details or tuning which are frequently introduced in Monte Carlo programs.  The $P_T$ distributions are generated entirely from known and universal DUPDFs.  For example, fits to $Z$ production data at the Tevatron using parton showers favour a large intrinsic partonic transverse momentum $\langle k_t \rangle \approx 2$ GeV, while confinement of partons inside the proton would imply a $\langle k_t \rangle \approx 0.3$ GeV \cite{primordial}.

The DGLAP-based parton showers used in \cite{Herwig,Pythia} are theoretically justified only in the limit of strongly-ordered transverse momentum, since only the collinear divergent part of the squared matrix element is kept in each parton branching.  Similarly, the CCFM-based parton shower used in \cite{Cascade} is strictly justified only in the limit of strongly-ordered angles, which reduces to the limit of strongly-ordered transverse momentum as long as $x$ is not too small.  In this limit, the transverse momentum generated in all evolution steps prior to the last is negligible.  Therefore, neglecting transverse momentum in every evolution step prior to the last should be a good approximation to the parton shower algorithms in which finite transverse momentum is generated at \emph{every} evolution step.

The goal of the present paper is to demonstrate that the $P_T$ distributions of electroweak bosons can be successfully generated by DUPDFs.  We do not aim to produce a \emph{better} description of the data than existing calculations, but rather a \emph{simpler} analytic description which reproduces the main features.  With this approach, it is easy to see the physical origin of the $P_T$ distributions and to identify the most important Feynman diagrams.  Since the DUPDFs are universal---that is, they apply equally well to all hard hadronic processes---it is important to check them in a new kinematic domain.

One topical application is the prediction of the cross section for diffractive Higgs boson production at the LHC \cite{Khoze:2001xm}, which is driven by the unintegrated\footnote{To be precise, the \emph{skewed} unintegrated gluon distribution is required.  However, in the relevant small-$x$ domain the skewed effect can be included by the Shuvaev prescription \cite{Shuvaev}.} gluon distribution $f_g(x,k_t^2,\mu^2)$, where $\mu$ is the hard scale of the subprocess.  At the moment, the only possibility to check the behaviour of UPDFs in the domain $k_t\ll \mu$ is to compare predictions with the observed $P_T$ distributions of $W$ and $Z$ bosons produced at the Tevatron.  We will show that the doubly-unintegrated quark distributions, generated directly from the known integrated PDFs under the `last-step' prescription, satisfactorily describe these data, including the region of interest, $P_T\ll M_{W,Z}$.

In Section \ref{sec:zkthad} we describe the formalism for $(z,k_t)$-factorisation at hadron-hadron colliders, and in Section \ref{sec:apply} we apply it to calculate the $P_T$ distributions of electroweak bosons.  Numerical results are given in Section \ref{sec:numerical}.

\section{$(z,k_t)$-factorisation at hadron-hadron colliders} \label{sec:zkthad}

\begin{figure}
  (a)\hfill (b)\hspace*{0.4\textwidth} \\
  \centering
  \begin{minipage}[c]{0.5\textwidth}
    \centering \includegraphics[width=0.8\textwidth]{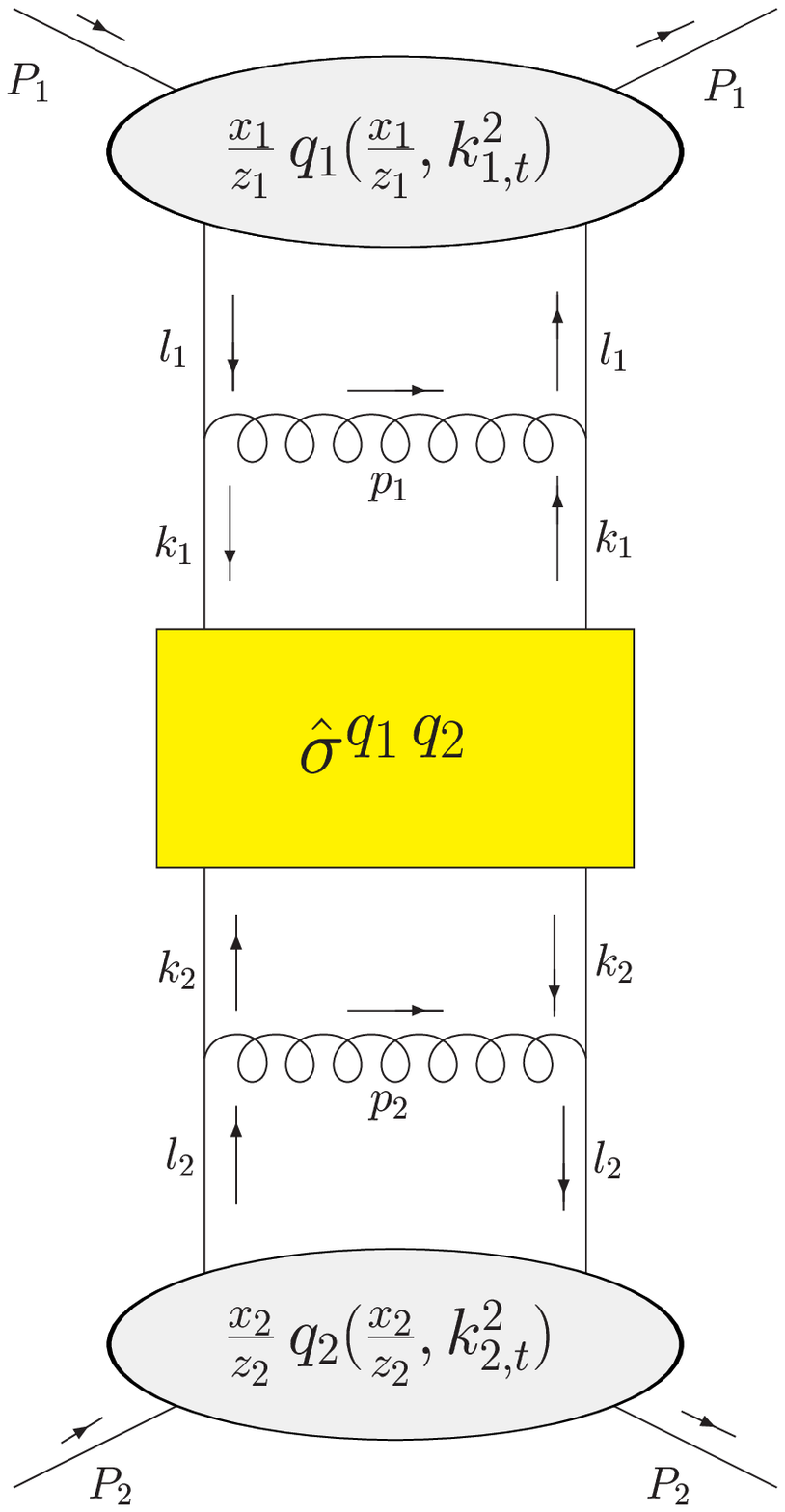}
  \end{minipage}%
  \begin{minipage}[c]{0.5\textwidth}
    \centering \includegraphics[width=0.8\textwidth]{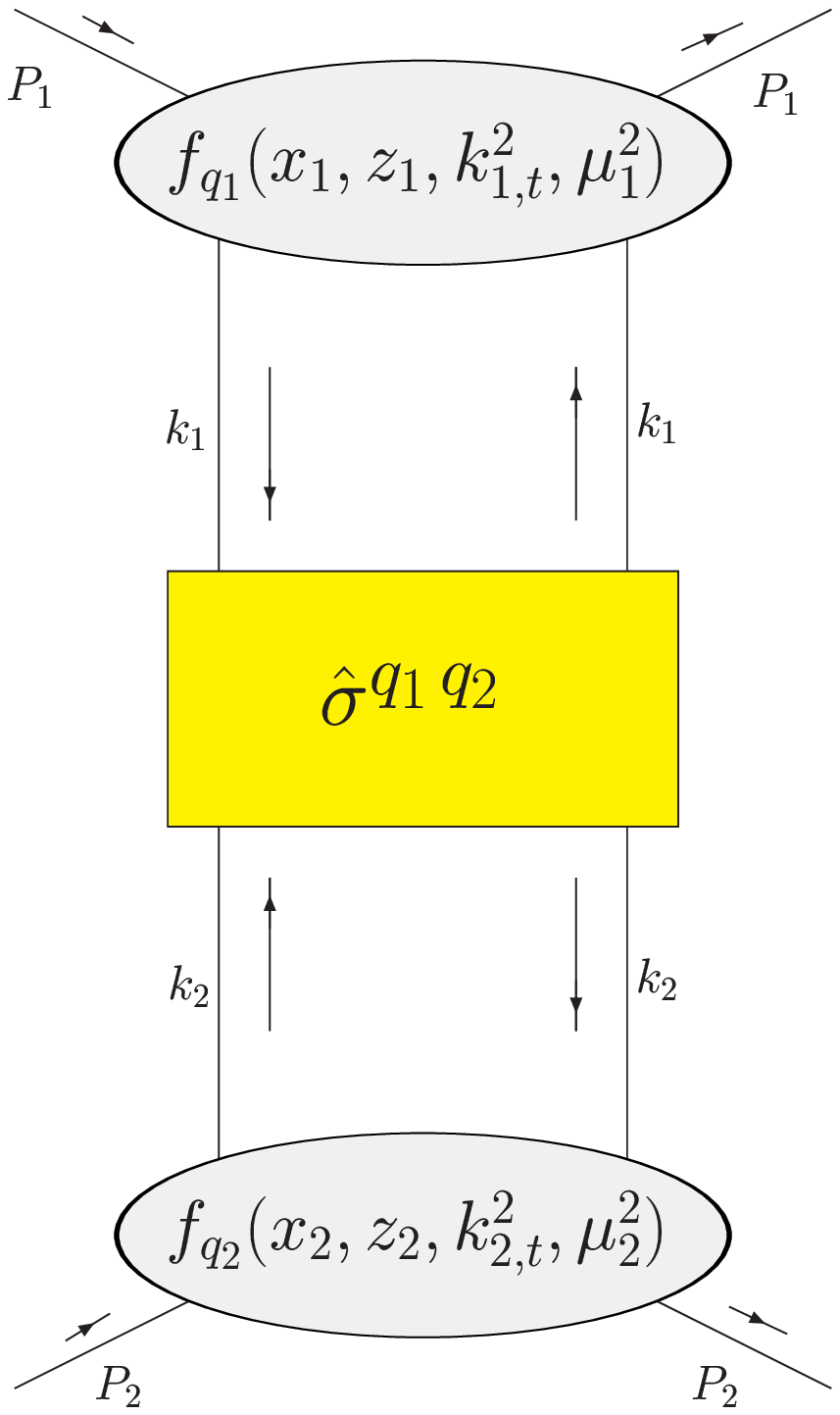}
  \end{minipage}
  \caption{(a) The transverse momentum of each parton entering the subprocess is generated by a single parton emission in the last evolution step.  (b) Illustration of $(z,k_t)$-factorisation:  the last evolution step is factorised into $f_{q_i}(x_i,z_i,k_{i,t}^2,\mu_i^2)$, where $i=1,2$.}
  \label{fig:zkt}
\end{figure}

We now extend the formalism of \cite{Watt}, which concerned deep-inelastic scattering (DIS), to hadron-hadron collisions.  The basic idea is illustrated in Fig.~\ref{fig:zkt}(a), which shows only one of the possible configurations.  All permutations of quarks and gluons must be included.  The arrows show the direction of the labelled momenta.  The blobs represent the familiar integrated PDFs.  The transverse momenta of the two incoming partons to the subprocess, represented by the rectangles labelled $\hat{\sigma}^{q_1\,q_2}$ in Fig.~\ref{fig:zkt}, are generated by a single parton emission in the last evolution step.

We use a Sudakov decomposition of the momenta of the two incoming partons:
\begin{equation} \label{eq:defki}
  k_i = x_i\,P_i - \beta_i\,P_j + {k_i}_\perp,
\end{equation}
where $(i,j)=(1,2)$ or $(2,1)$.  We work in the centre-of-mass frame of the colliding hadrons and neglect the hadron masses so that the squared centre-of-mass energy is $s\equiv(P_1+P_2)^2\simeq 2\,P_1\cdot P_2$.  Then,\footnote{The `plus' and `minus' light-cone components of a 4-momentum $P$ are $P^{\pm}\equiv P^0 \pm P^3$.}
\begin{equation}
  P_1 = (P_1^+,P_1^-,{P_1}_\perp) = \sqrt{s}\,(1,0,\vec{0}), \qquad P_2 = \sqrt{s}\,(0,1,\vec{0}) \qquad {k_i}_\perp = (0,0,\vec{k_{i,t}}).
\end{equation}
The penultimate propagators in the evolution ladder have momenta $l_i=x_iP_i/z_i$, so that the partons emitted in the last step have momenta
\begin{equation}
  p_i = l_i - k_i = \frac{x_i}{z_i}\,(1-z_i)\,P_i + \beta_i\,P_j - {k_i}_\perp.
\end{equation}
The on-shell condition for the emitted partons, $p_i^2 = 0$, determines
\begin{equation} \label{eq:betai}
  \beta_i = \frac{z_i\,r_i}{x_i\,(1-z_i)},
\end{equation}
where $r_i\equiv k_{i,t}^2/s$, so that the two incoming partons have virtuality $k_i^2 = -k_{i,t}^2/(1-z_i)$.  The total momentum going into the subprocess labelled $\hat{\sigma}^{q_1\,q_2}$ in Fig.~\ref{fig:zkt} is
\begin{equation}
  q \equiv k_1+k_2 = (x_1-\beta_2)\,P_1 + (x_2-\beta_1)\,P_2 + q_\perp,
\end{equation}
where $q_\perp = {k_1}_\perp + {k_2}_\perp$.  The kinematic variables obey the ordering
\begin{equation} \label{eq:kinord}
  0 < \beta_j < x_i < z_i < 1.
\end{equation}

If $p_i$ are gluon momenta, then we must additionally impose angular-ordering\footnote{``Angular ordering'' is in fact a misnomer.  It is rapidity ordering which should be applied.} constraints due to colour coherence \cite{CCFM}:
\begin{equation}
  \xi_1 < \Xi < \xi_2,
\end{equation}
where $\xi_i \equiv p_i^- / p_i^+$ and $\Xi \equiv q^-/q^+$.  That is, the subprocess separates gluons emitted from each of the two hadrons.  This condition leads to a suppression of soft gluon emission:
\begin{equation} \label{eq:angord}
  z_i < \frac{\mu_i}{\mu_i+k_{i,t}},
\end{equation}
with $\mu_1 \equiv x_1\sqrt{s\,\Xi}$ and $\mu_2 \equiv x_2\sqrt{s\,/\,\Xi}$.

The partonic cross sections, represented by the rectangles in Fig.~\ref{fig:zkt}, depend on $x_i$, $z_i$, and $k_{i,t}$ through the momenta of the incoming partons $k_i$ \eqref{eq:defki}. Therefore, the partonic cross sections must be convoluted with parton distributions depending on the same variables.  We can define DUPDFs $f_a(x,z,k_t^2,\mu^2)$ which satisfy the normalisation conditions
\begin{equation} \label{eq:norm}
  \int_x^1\!\dif{z}\int_0^{\mu^2}\!\diff{k_t^2}\,f_a(x,z,k_t^2,\mu^2)=a(x,\mu^2),
\end{equation}
for fixed $x$ and $\mu$, independent of $z$ and $k_t$.  Here, $a(x,\mu^2)=x\,g(x,\mu^2)$ or $x\,q(x,\mu^2)$ are the conventional integrated PDFs.  Explicit formulae for the DUPDFs are given in \cite{Watt}.  The doubly-unintegrated quark distribution is
\begin{equation}
  f_q(x,z,k_t^2,\mu^2) = T_q(k_t^2,\mu^2)\,\frac{\alpha_S(k_t^2)}{2\pi}\;\left[\,P_{qq}(z)\,\frac{x}{z}q\left(\frac{x}{z},k_t^2\right)\,\Theta\left(\frac{\mu}{\mu+k_t}-z\right) + P_{qg}(z)\,\frac{x}{z}g\left(\frac{x}{z},k_t^2\right)\,\right],
\end{equation}
where $P_{qq}(z)$ and $P_{qg}(z)$ are the unregulated LO DGLAP splitting kernels, and the quark Sudakov form factor is
\begin{equation} \label{eq:qSud}
  T_q(k_t^2,\mu^2) = \exp\left(-\int_{k_t^2}^{\mu^2}\!\diff{\kappa_t^2}\,\frac{\alpha_S(\kappa_t^2)}{2\pi}\,\int_0^{\frac{\mu}{\mu+\kappa_t}}\!\dif{\zeta }\,P_{qq}(\zeta )\right).
\end{equation}
The doubly-unintegrated gluon distribution is
\begin{equation}
  f_g(x,z,k_t^2,\mu^2) = T_g(k_t^2,\mu^2)\,\frac{\alpha_S(k_t^2)}{2\pi}\;\left[\sum_q P_{gq}(z)\,\frac{x}{z}q\left(\frac{x}{z},k_t^2\right) + P_{gg}(z)\,\frac{x}{z}g\left(\frac{x}{z},k_t^2\right)\,\Theta\left(\frac{\mu}{\mu+k_t}-z\right)\,\right],
\end{equation}
where the gluon Sudakov form factor is
\begin{equation} \label{eq:gSud}
  T_g(k_t^2,\mu^2) = \exp\left[-\int_{k_t^2}^{\mu^2}\!\diff{\kappa_t^2}\,\frac{\alpha_S(\kappa_t^2)}{2\pi}\,\left( \int_{\frac{\kappa_t}{\mu+\kappa_t}}^{\frac{\mu}{\mu+\kappa_t}}\!\dif{\zeta }\;\zeta \,P_{gg}(\zeta ) + n_F\,\int_0^1\!\dif{\zeta}\,P_{qg}(\zeta)\right)\right],
\end{equation}
with $n_F$ the number of active flavours.

In terms of the DUPDFs, the hadronic cross section $\sigma$ is related to the partonic cross sections $\hat{\sigma}^{a_1\,a_2}$ by the $(z,k_t)$-factorisation formula
\begin{equation} \label{eq:zkt}
  \sigma = \sum_{a_1,\,a_2}\int_0^1\!\diff{x_1}\!\int_0^1\!\diff{x_2}\!\int_{x_1}^1\!\dif{z_1}\!\int_{x_2}^1\!\dif{z_2}\!\int_0^{\infty}\!\diff{k_{1,t}^2}\!\int_0^{\infty}\!\diff{k_{2,t}^2}\;f_{a_1}(x_1,z_1,k_{1,t}^2,\mu_1^2)\;f_{a_2}(x_2,z_2,k_{2,t}^2,\mu_2^2)\;\hat{\sigma}^{a_1\,a_2}.
\end{equation}
This formula is represented schematically in Fig.~\ref{fig:zkt}(b) for the case where the partons $a_1$ and $a_2$ are both quarks.  The partonic cross sections in \eqref{eq:zkt} are given by
\begin{equation}
  \dif\hat{\sigma}^{a_1\,a_2} = \dif\Phi^{a_1\,a_2}\,\lvert\mathcal{M}^{a_1\,a_2}\rvert^2 \,/\, F^{a_1\,a_2},
\end{equation}
where the flux factor $F^{a_1\,a_2} = 2\,x_1\,x_2\,s$.  The last evolution steps in Fig.~\ref{fig:zkt}(a) factorise from the rest of the diagram, to give the LO DGLAP splitting kernels, in the leading logarithmic approximation (LLA) where only the $\dif{k_{i,t}^2}/k_{i,t}^2$ term is kept.  Therefore, $\lvert\mathcal{M}^{a_1\,a_2}\rvert^2$ is calculated with the replacement $k_i \to x_i\,P_i$ in the numerator in order to keep only this collinear divergent term.  However, any propagator virtualities appearing in the denominator of $\lvert\mathcal{M}^{a_1\,a_2}\rvert^2$ may be evaluated with the full kinematics, as may the phase space element $\dif\Phi^{a_1\,a_2}$.

For this approach to work, it is vital that the $\dif{k_{i,t}^2}/k_{i,t}^2$ term is obtained \emph{only} from ladder-type diagrams like that in Fig.~\ref{fig:zkt}(a), and not from interference (non-ladder) diagrams.  This is true if we use a physical gauge for the gluon, where only the two transverse polarisations propagate.  For hadron-hadron collisions, the natural choice is the planar gauge where the sum over gluon polarisations is performed using 
\begin{equation}
  \label{eq:planar}
  d_{\mu\nu}(k) = -g_{\mu\nu} + \frac{k_\mu\,n_\nu + n_\mu\,k_\nu}{k\cdot n},
\end{equation}
where we take the gauge-fixing vector $n=x_1\,P_1+x_2\,P_2$.  Such a gauge choice ensures that the $\dif{k_{i,t}^2}/k_{i,t}^2$ term is obtained from ladder-type diagrams on \emph{both} sides of the subprocess represented by the rectangle in Fig.~\ref{fig:zkt}(a) \cite{Dokshitzer:hw}.

A related requirement is that terms beyond the leading $\dif{k_{i,t}^2}/k_{i,t}^2$ term, coming from non-ladder diagrams, for example, give a negligible contribution.  Such terms are proportional to the Sudakov variable $\beta_i$ \eqref{eq:betai} and hence vanish in the limit that $z_i\to 0$ or $k_{i,t}^2/s\to 0$ (unless $z_i\to 1$).  Away from these limits it is not obvious that these ``beyond LLA'' terms will be small, a necessary condition for the factorisation to hold.  For the case of inclusive jet production in DIS and working in an axial gluon gauge, it was observed in \cite{Watt} that the main effect of the extra terms was to suppress soft gluon emission.  When the angular-ordering constraint \eqref{eq:angord} was applied, the extra terms were found to make a negligible difference to the cross section.  For hadron-hadron collisions, although the number of possible non-ladder diagrams is larger, it is therefore reasonable to expect that the extra terms will have little numerical effect, at least for $k_{i,t}$ less than the hard scale of the subprocess.  A similar argument is made to justify the approximation made in the DGLAP-based parton showers used in Monte Carlo simulations, where only the collinear divergent part of the squared matrix element for each parton branching is kept and angular ordering is imposed in all evolution steps to account for some of the missing terms.  Here, we are more conservative and apply this approximation to the \emph{last} evolution step only.

The DUPDFs in \eqref{eq:zkt} are only defined for $k_{i,t}>\mu_0$, where $\mu_0\sim 1$ GeV is the minimum scale for which DGLAP evolution of the integrated PDFs is valid.  The approximation of the $k_{i,t}<\mu_0$ contribution made in \cite{Watt} was to take the limit $k_{i,t}\to 0$ in the kinematic variables (and in $\hat{\sigma}^{a_1\,a_2}$), then to make the replacement
\begin{equation} \label{eq:lowktnorm}
  \int_{x_i}^1\!\dif{z_i}\int_0^{\mu_0^2}\!\diff{k_{i,t}^2}\,f_{a_i}(x_i,z_i,k_{i,t}^2,\mu_i^2) = a_i(x_i,\mu_0^2)\,T_{a_i}(\mu_0^2,\mu_i^2),
\end{equation}
where $T_{a_i}(\mu_0^2,\mu_i^2)$ are the Sudakov form factors \eqref{eq:qSud} or \eqref{eq:gSud}.  This replacement ensures that the normalisation conditions \eqref{eq:norm} are satisfied (see \cite{Watt}).  A better approximation, which retains the $k_{i,t}$ dependence, is to take the limit $z_i\to 0$ in the kinematic variables, then to make the replacement
\begin{equation} \label{eq:lowkt}
  \int_{x_i}^1\!\dif{z_i}\,f_{a_i}(x_i,z_i,k_{i,t}^2,\mu_i^2) \equiv f_{a_i}(x_i,k_{i,t}^2,\mu_i^2) = \frac{k_{i,t}^2}{\mu_0^2}\,a_i(x_i,\mu_0^2)\,T_{a_i}(\mu_0^2,\mu_i^2),
\end{equation}
where we have used (23) of \cite{Watt}.  The requirement that $f_{a_i}(x_i,k_{i,t}^2,\mu_i^2)\sim k_{i,t}^2$ as $k_{i,t}^2\to 0$ is a consequence of gauge invariance \cite{GribovAskew}.

A more complicated extrapolation of the DUPDFs for $k_t<\mu_0$, which allows both the $k_t$ and $z$ dependence to be retained in the kinematic variables, is to assume the polynomial form
\begin{equation}
  \label{eq:newlowkt}
  f_a(x,z,k_t^2,\mu^2) = \frac{k_t^2}{\mu_0^2}\left[A(x,z,\mu^2) + \frac{k_t^2}{\mu_0^2}\,B(x,z,\mu^2)\right].
\end{equation}
The two coefficients $A$ and $B$ can be determined to ensure continuity at $k_t = \mu_0$ and the correct normalisation \eqref{eq:lowktnorm}, leading to
\begin{align}
  A(x,z,\mu^2) &= -f_a(x,z,\mu_0^2,\mu^2) + 2\,a(x,\mu_0^2)\,T_a(\mu_0^2,\mu^2)\,/\,(1-x), \\
  B(x,z,\mu^2) &= 2\,f_a(x,z,\mu_0^2,\mu^2) - 2\,a(x,\mu_0^2)\,T_a(\mu_0^2,\mu^2)\,/\,(1-x).
\end{align}
If \eqref{eq:newlowkt} becomes negative, which is possible for $k_t\ll\mu_0$, then $f_a(x,z,k_t^2,\mu^2)$ is simply set to zero.

This extrapolation of the perturbative formulae accounts for some non-perturbative ``intrinsic'' $k_t$ of the initial partons, which is often parameterised by a Gaussian distribution with $\langle k_t \rangle \lesssim \mu_0$.  Numerical results are insensitive to the precise form \eqref{eq:lowktnorm}, \eqref{eq:lowkt}, or \eqref{eq:newlowkt} used for the $k_t<\mu_0$ contribution.  However, we will use the form \eqref{eq:newlowkt} which ensures continuity at $k_t=\mu_0$.

\section{Application to the $P_T$ distribution of electroweak bosons} \label{sec:apply}

\begin{figure}
  \centering
    \includegraphics[width=0.5\textwidth,clip]{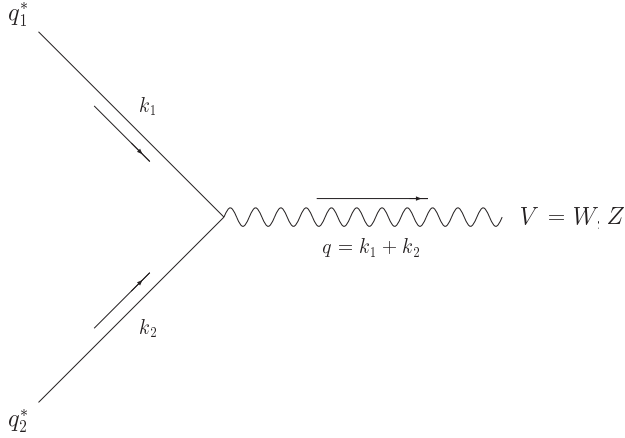}
  \caption{LO Feynman diagram contributing to the $P_T$ distribution of $W$ or $Z$ bosons in the $(z,k_t)$-factorisation approach.}
  \label{fig:qqV}
\end{figure}
\begin{figure}
  \centering
    \includegraphics[width=0.5\textwidth,clip]{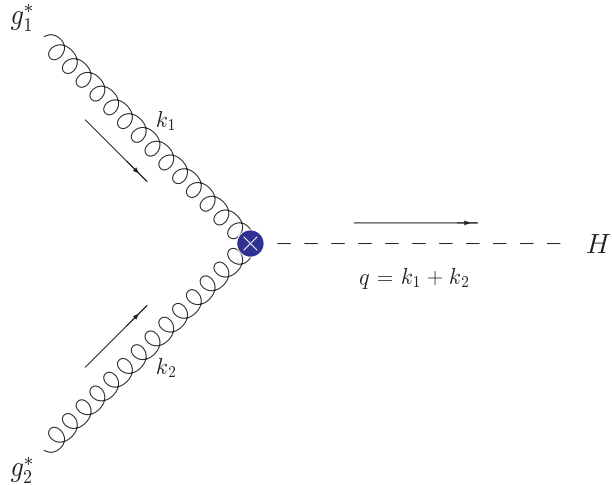}
  \caption{LO Feynman diagram contributing to the $P_T$ distribution of Higgs bosons in the $(z,k_t)$-factorisation approach.  The cross represents the effective vertex in the limit $M_H\ll 2m_t$.}
  \label{fig:ggH}
\end{figure}

Perhaps the simplest application of $(z,k_t)$-factorisation at hadron-hadron colliders is electroweak boson production, where at LO the subprocess is simply $q_1^*\,q_2^*\to V$, illustrated in Fig.~\ref{fig:qqV}, or $g_1^*\,g_2^*\to H$, illustrated in Fig.~\ref{fig:ggH}.  Here, the `$^*$' indicates that the incoming partons, with momenta $k_i$ given by \eqref{eq:defki}, are off-shell with virtuality $k_i^2=-k_{i,t}^2/(1-z_i)$.  In the collinear approximation, these diagrams give the Born level estimate of the total cross section $\sigma$.  When each parton carries finite transverse momentum $\vec{k_{i,t}}$, the final electroweak boson has transverse momentum $\vec{q_t} = \vec{k_{1,t}}+\vec{k_{2,t}}$, so we can calculate the $P_T$ distribution
\begin{equation}
  \frac{\dif\sigma}{\dif P_T} = \sigma\;\delta(q_t-P_T),
\end{equation}
with $\sigma$ given by the $(z,k_t)$-factorisation formula \eqref{eq:zkt}.

The partonic differential cross sections for $W$ or $Z$ production are
\begin{equation} \label{eq:WZPThat}
  \frac{\dif\hat{\sigma}}{\dif P_T}(q_1^*\,q_2^*\to V) = \frac{\pi}{N_C}\,\sqrt{2}\,G_F\,M_V^2\,V_V^2\;\delta(q^2-M_V^2)\;\delta(q_t-P_T),
\end{equation}
where $N_C=3$ is the number of colours, $G_F$ is the Fermi coupling constant, $V_W^2 \equiv \lvert V_{q_1q_2}\rvert^2$ is the CKM matrix element squared, and $V_Z^2 \equiv V_q^2 + A_q^2$ is the sum of the vector and axial vector couplings squared.

The dominant mechanism for SM Higgs production in hadron-hadron collisions is by gluon-gluon fusion via a top quark loop.  For the case where $M_H \ll 2m_t$, the well-known effective $ggH$ vertex can be derived from the Lagrangian \cite{ggHvertex}
\begin{equation}
  \label{eq:LggH}
  \mathcal{L}_{\mathrm{eff}} = -\frac{1}{4}\left(1-\frac{\alpha_S(M_H^2)}{3\pi}\frac{H}{v}\right)\,G^a_{\mu\nu}\,G^{a\,\mu\nu},
\end{equation}
where $v^2=(\sqrt{2}\,G_F)^{-1}$, $G^a_{\mu\nu}$ is the gluon field strength tensor, and $H$ is the Higgs field.  The partonic differential cross section for SM Higgs production is then
\begin{equation} \label{eq:HPThat}
  \frac{\dif\hat{\sigma}}{\dif P_T}(g_1^*\,g_2^*\to H) = \frac{\sqrt{2}\,G_F\,x_1\,x_2\,s}{576\,\pi}\,\alpha_S^2(M_H^2)\;\delta(q^2-M_H^2)\;\delta(q_t-P_T).
\end{equation}

For $q_t=0$, these expressions \eqref{eq:WZPThat} and \eqref{eq:HPThat} are exactly as in the collinear approximation.  The difference arises when we consider the precise kinematics
\begin{gather}
  q^2 = s\,\left[(x_1-\beta_2)(x_2-\beta_1)-R\right],\qquad R\equiv q_t^2/s, \\
  q_t^2 = \lvert\vec{k_{1,t}}+\vec{k_{2,t}}\rvert^2 = k_{1,t}^2 + k_{2,t}^2 + 2\,k_{1,t}\,k_{2,t}\,\cos\phi.
\end{gather}

Applying the $(z,k_t)$-factorisation formula \eqref{eq:zkt}, the first delta function in \eqref{eq:WZPThat} and \eqref{eq:HPThat} can be used to do the $x_2$ integration in \eqref{eq:zkt}, while the second delta function can be used to do the $k_{2,t}$ integration.  In addition, we need to average over the azimuthal angle $\phi$ between $\vec{k_{1,t}}$ and $\vec{k_{2,t}}$.

The final hadronic differential cross sections for $W$ or $Z$ production are
\begin{multline} \label{eq:WZPT}
  \frac{\dif\sigma}{\dif P_T} = \frac{\pi}{N_C}\,\sqrt{2}\,G_F\,\tau\sum_{x_2=x_2^\pm}\sum_{k_{2,t}=k_{2,t}^\pm}\int_0^1\!\diff{x_1}\!\int_{x_1}^1\!\dif{z_1}\!\int_{x_2}^1\!\dif{z_2}\!\int_{0}^{\infty}\!\diff{k_{1,t}^2}\!\int_0^{2\pi}\!\frac{\dif \phi}{2\pi}\;\frac{2\,P_T\,\Theta(k_{2,t})}{k_{2,t}\lvert k_{2,t}+k_{1,t}\cos\phi\rvert}\\\times \frac{1}{x_1\,x_2-\beta_1\,\beta_2}\;\sum_{q_1,\,q_2} V_V^2\;f_{q_1}(x_1,z_1,k_{1,t}^2,\mu_1^2)\;f_{q_2}(x_2,z_2,k_{2,t}^2,\mu_2^2),
\end{multline}
with $\tau\equiv M_V^2/s$, $k_{2,t}^\pm\equiv -k_{1,t}\,\cos\phi \pm \sqrt{P_T^2 -k_{1,t}^2\sin^2\phi}$, and
\begin{equation}
  x_2^\pm \equiv \frac{1}{2x_1}\left\{\tau+R+x_1\beta_1+\frac{z_2r_2}{1-z_2}\pm \sqrt{\left(\tau+R+x_1\beta_1+\frac{z_2r_2}{1-z_2}\right)^2 - 4x_1\beta_1\frac{z_2r_2}{1-z_2}}\right\}.
\end{equation}
In practice, the kinematic constraints \eqref{eq:kinord} mean that the $x_2=x_2^-$ solution does not contribute.  The corresponding result for SM Higgs production is
\begin{multline} \label{eq:HPT}
  \frac{\dif\sigma}{\dif P_T} = \frac{\sqrt{2}\,G_F}{576\,\pi}\,\alpha_S^2(M_H^2)\sum_{x_2=x_2^\pm}\sum_{k_{2,t}=k_{2,t}^\pm}\int_0^1\!\diff{x_1}\!\int_{x_1}^1\!\dif{z_1}\!\int_{x_2}^1\!\dif{z_2}\!\int_{0}^{\infty}\!\diff{k_{1,t}^2}\!\int_0^{2\pi}\!\frac{\dif \phi}{2\pi}\;\frac{2\,P_T\,\Theta(k_{2,t})}{k_{2,t}\lvert k_{2,t}+k_{1,t}\cos\phi\rvert}\\\times \frac{x_1\,x_2}{x_1\,x_2-\beta_1\,\beta_2}\;f_g(x_1,z_1,k_{1,t}^2,\mu_1^2)\;f_g(x_2,z_2,k_{2,t}^2,\mu_2^2),
\end{multline}
where $\tau\equiv M_H^2/s$ and $k_{2,t}^\pm$ and $x_2^\pm$ are as above.  Note that we have taken the $ggH$ vertex in the $M_H\ll 2m_t$ limit.  For $M_H < 2m_t$, the correction to the total cross section due to the top quark mass can be approximated \cite{Ellis:qj} by a factor
\begin{equation}
  \label{eq:topcorr}
  \left[1+\left(\frac{M_H}{2m_t}\right)^2\right]^2.
\end{equation}
The $k_{i,t}<\mu_0$ contributions of \eqref{eq:WZPT} and  \eqref{eq:HPT} are accounted for using the approximation \eqref{eq:newlowkt}.\footnote{Other contributions in the region $k_{i,t}<\mu_0$, such as the inclusion of additional intrinsic partonic transverse momentum with $\langle k_t \rangle \approx 0.3$ GeV, would only affect the $P_T$ distributions at very low $P_T\lesssim 1$ GeV.}

\section{The $K$-factor}

In the collinear approximation, higher order QCD corrections to the LO diagrams, $q_1\,q_2\to V$ or $g_1\,g_2\to H$, are known to be significant when calculating the total cross section.  The ratio of the corrected result to the leading order result is the so-called $K$-factor.  A part of these higher order corrections is kinematic in nature, arising from real parton emission, which we have already accounted for at LO in the $(z,k_t)$-factorisation approach (see Fig.~\ref{fig:zkt}).  Another part comes from the logarithmic loop corrections which have already been included in the Sudakov form factors \eqref{eq:qSud} and \eqref{eq:gSud}.  However, we need to include the non-logarithmic loop corrections arising, for example, from the gluon vertex correction to Figs.~\ref{fig:qqV} and ~\ref{fig:ggH}.

A large part of these non-logarithmic corrections have a semi-classical nature and may be obtained from the analytic continuation of the double logarithm in the Sudakov form factors in going from spacelike (DIS) to timelike (Drell-Yan) electroweak boson momenta \cite{Kfactor}.  Accounting for the running of the strong coupling $\alpha_S$, the Sudakov form factors in the DLLA are
\begin{equation}
  \label{eq:qSudDLLA}
   T_q(k_t^2,\mu^2) = \exp\left(-\frac{C_F}{2\,\pi\,b}\,L\ln L\right)
\end{equation}
and
\begin{equation}
  \label{eq:gSudDLLA}
  T_g(k_t^2,\mu^2) = \exp\left(-\frac{C_A}{2\,\pi\,b}\,L\ln L\right),
\end{equation}
where $L\equiv \ln(\mu^2/\Lambda_{\textrm{QCD}}^2)$, $b=(33-2n_F)/(12\pi)$, and the colour factors are $C_F=4/3$ and $C_A=3$.

Replacing $\mu^2$ by $-\mu^2$, we obtain the $\pi^2$-enhanced part of the $K$-factors:
\begin{equation} \label{eq:KqqV}
  K(q_1^*\,q_2^*\to V) \simeq \left\lvert \frac{T_q(k_t^2,-\mu^2)}{T_q(k_t^2,\mu^2)}\right\rvert^2 \simeq \exp\left(C_F\,\frac{\alpha_S(\mu^2)}{2\,\pi}\,\pi^2\right)
\end{equation}
and
\begin{equation} \label{eq:KggH}
  K(g_1^*\,g_2^*\to H) \simeq \left\lvert \frac{T_g(k_t^2,-\mu^2)}{T_g(k_t^2,\mu^2)}\right\rvert^2 \simeq \exp\left(C_A\,\frac{\alpha_S(\mu^2)}{2\,\pi}\,\pi^2\right).
\end{equation}
A particular scale choice $\mu^2 = P_T^{4/3}\,M_{V,H}^{2/3}$ has been found \cite{Kulesza:1999gm} to eliminate certain sub-leading logarithms in the Sudakov form factors.  Therefore, we choose this scale to evaluate $\alpha_S(\mu^2)$ in \eqref{eq:KqqV} and \eqref{eq:KggH}.

\section{Numerical results} \label{sec:numerical}
\subsection{$W$ and $Z$ boson production at the Tevatron}
\begin{figure}
  \centering
  \includegraphics[width=\textwidth,clip]{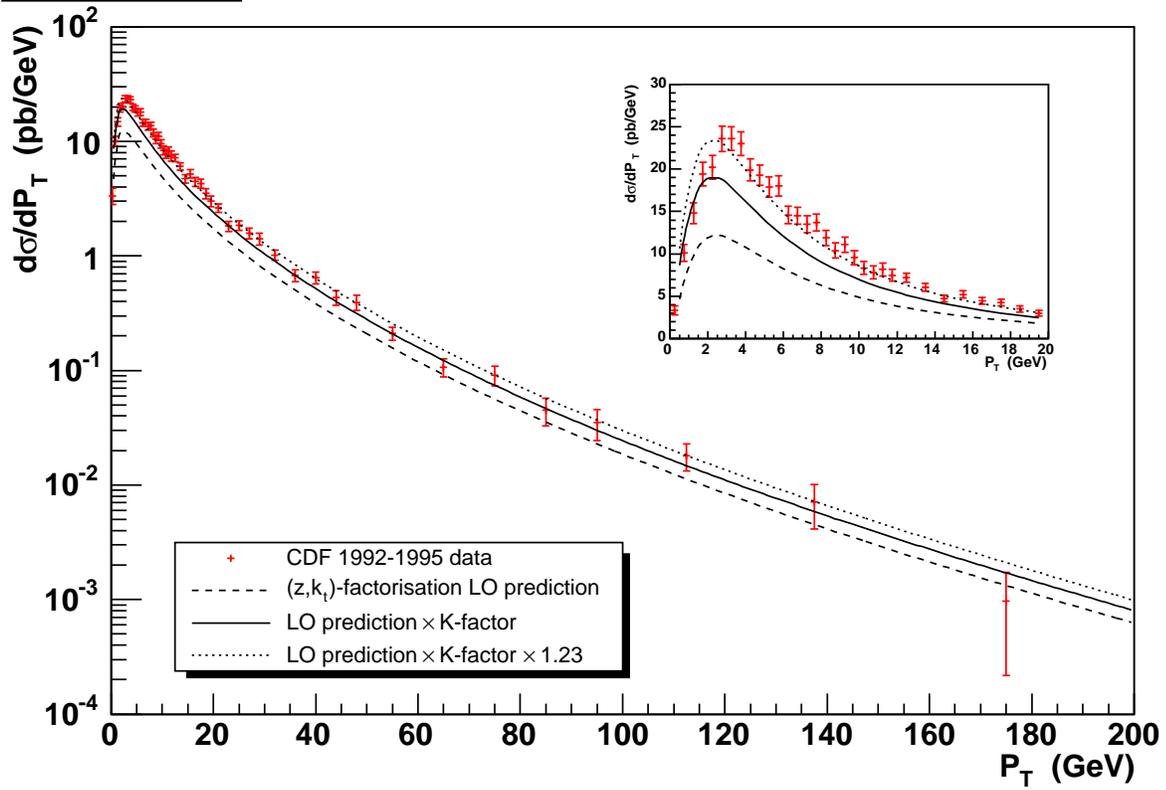}
  \caption{$P_T$ distribution of $Z$ bosons compared to CDF data \cite{CDFZ}.}
  \label{fig:CDFZ}
\end{figure}

\begin{figure}
  \centering
  \includegraphics[width=\textwidth,clip]{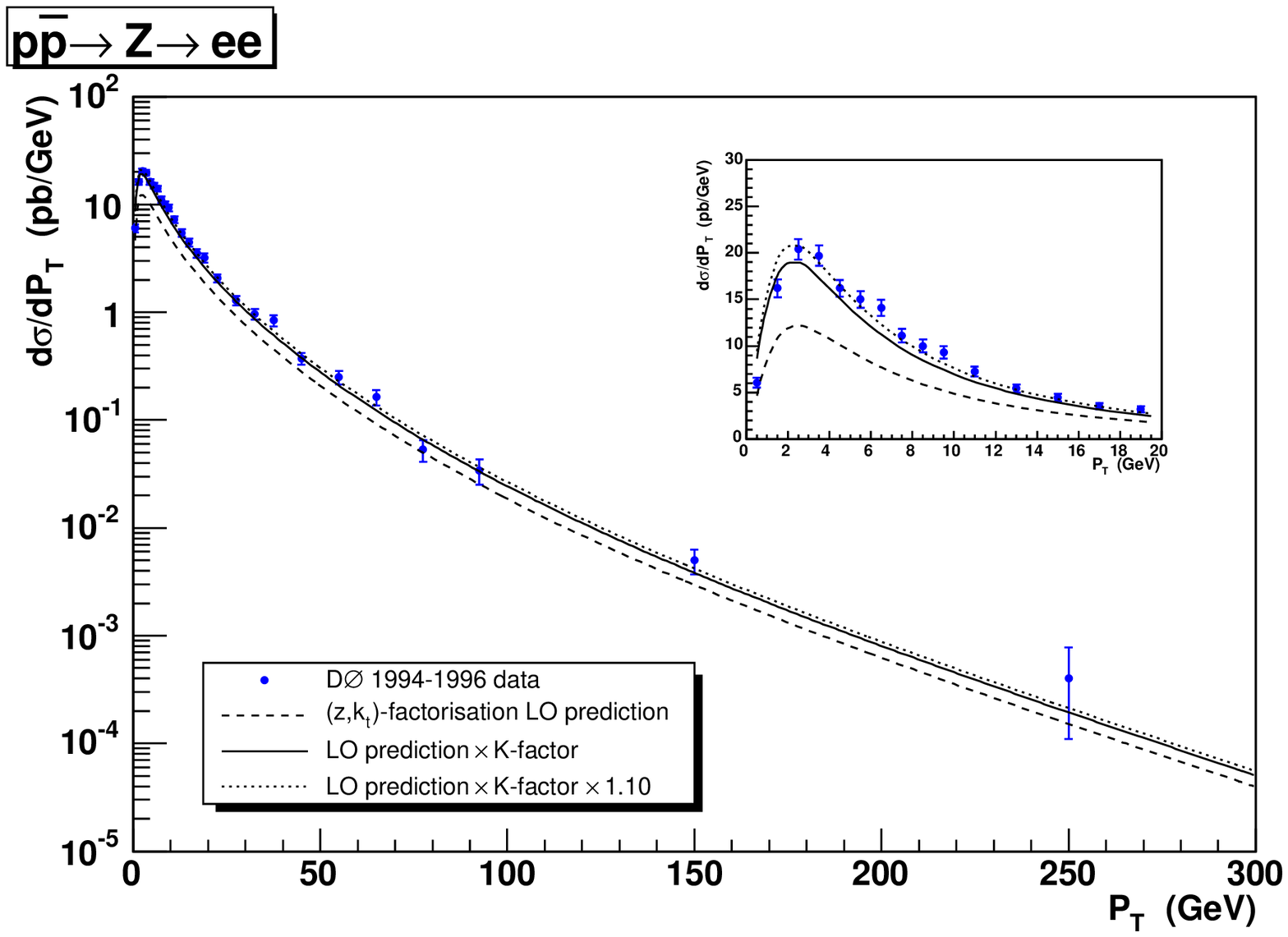}
  \caption{$P_T$ distribution of $Z$ bosons compared to D{\O} data \cite{D0Z}.}
  \label{fig:D0Z}
\end{figure}

\begin{figure}
  \centering
  \includegraphics[width=\textwidth,clip]{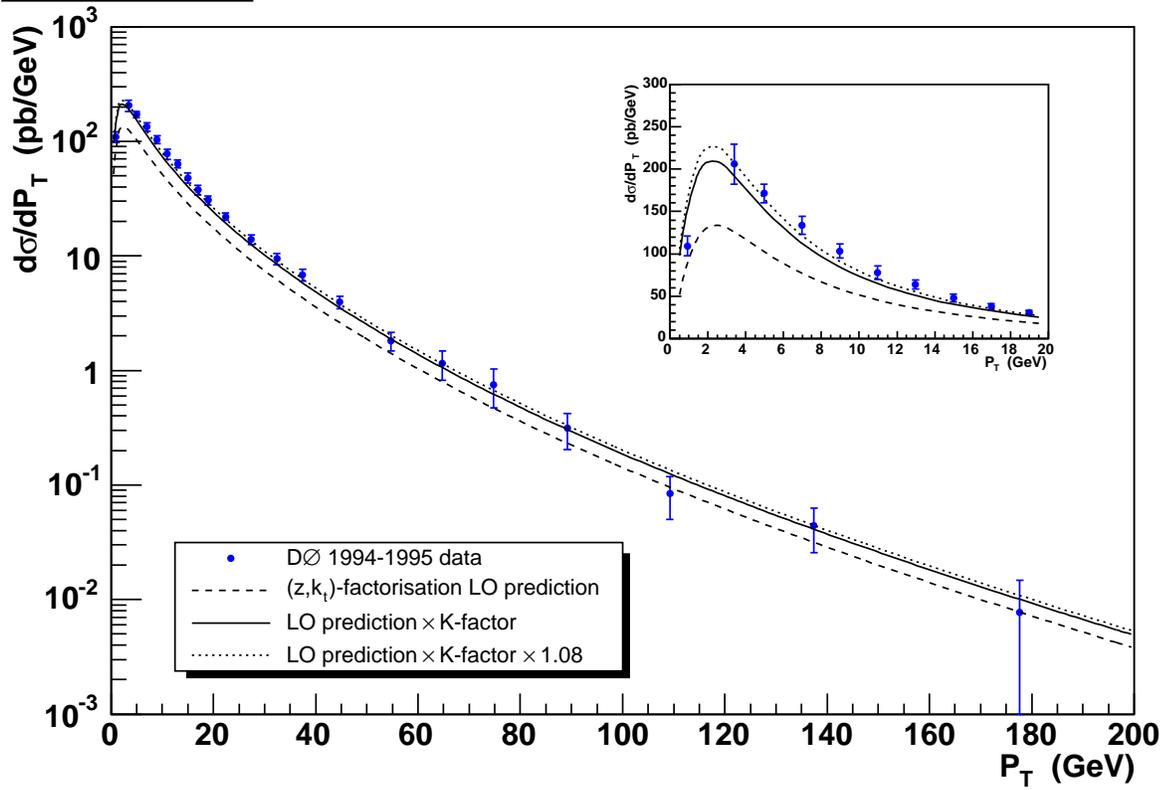}
  \caption{$P_T$ distribution of $W$ bosons compared to D{\O} data \cite{D0W}.}
  \label{fig:D0W}
\end{figure}

The $P_T$ distributions of produced $W$ and $Z$ bosons were measured by the CDF \cite{CDFZ} and D{\O} \cite{D0Z,D0W} Collaborations during the Tevatron Run 1, in $p\bar{p}$ collisions at a centre-of-mass energy of $\sqrt{s}=1.8$ TeV.  Measurements were made of $W\to e\nu$ and $Z\to ee$ decays; therefore, we must multiply the theoretical predictions for $W$ or $Z$ production by the appropriate leptonic branching ratios.\footnote{BR$(W\to e\nu) = 0.1072$ and BR$(Z\to ee)=0.03363$ \cite{Hagiwara:fs}.}  We use the MRST2001 LO PDFs \cite{Martin:2002dr} as input, with $\mu_0^2=1.25$ GeV$^2$.  The LO predictions for the $P_T$ distributions \eqref{eq:WZPT}, integrated over bins of 1 GeV, are shown by the dashed lines in Figs.~\ref{fig:CDFZ}, \ref{fig:D0Z} and \ref{fig:D0W}.  The integrated luminosity uncertainty (3.9\% for CDF or 4.4\% for D{\O}) is not included in the error bars for the plotted data.  The LO predictions multiplied by the $K$-factor \eqref{eq:KqqV} are shown by the solid lines.  Although the $K$-factor makes up the major part of the discrepancy, we see that the solid lines in Figs.~\ref{fig:CDFZ}, \ref{fig:D0Z}, and \ref{fig:D0W} still underestimate the precise measurements at small $P_T$ to some extent.  Normalising to the total measured cross sections (248 pb, 221 pb, and 2310 pb) by factors 1.23, 1.10, and 1.08 respectively, as shown by the dotted lines in Figs.~\ref{fig:CDFZ}, \ref{fig:D0Z}, and \ref{fig:D0W}, gives a very good description of the data over the entire $P_T$ range.

Although it makes sense to take $\mu_0^2$ as low as possible, the insensitivity of the $P_T$ distributions to the $k_t<\mu_0$ treatment can be demonstrated by taking $\mu_0^2=2.5$ GeV$^2$.  The $P_T$ distributions obtained are practically identical to those with $\mu_0^2=1.25$ GeV$^2$, with less than a 0.5\% change in the total cross sections.

Notice that the predicted $P_T$ distribution of $Z$ bosons peaks about 0.5--1.0 GeV below the CDF data (Fig.~\ref{fig:CDFZ}).  One possible explanation for this is provided by non-perturbative power corrections, part of which may be interpreted as a negative correction of about $-3$ GeV$^2$ to the factorisation scale at which the integrated PDFs are evaluated \cite{Guffanti:2000ep}.  Such a shift in the factorisation scale is found to move the peak of the $P_T$ distribution about 0.2 GeV in the direction of the CDF data, with a slightly larger normalisation factor of 1.24.


The small residual discrepancy between the solid lines in Figs.~\ref{fig:CDFZ}, \ref{fig:D0Z}, and \ref{fig:D0W} and the data is easily  understood.  Note that the MRST2001 LO PDFs \cite{Martin:2002dr} have been determined by a global fit to data using the conventional collinear approximation.  A more precise treatment would fit the integrated PDFs, used as input to the last evolution step, to the proton structure function $F_2$, for example, using the $(z,k_t)$-factorisation formalism at LO.  We would expect this treatment to give slightly larger integrated PDFs than the conventional sets by a factor of $\approx 1.1$ and so eliminate the small discrepancy between the $(z,k_t)$-factorisation predictions and the data.

Alternatively, it was found in \cite{Watt} that the major higher order corrections to the inclusive jet cross section in DIS could be accounted for by adding extra parton emissions to the LO diagram, $\gamma^*q^*\to q$.  In an axial gluon gauge, ladder-type diagrams gave the dominant contributions.  Thus, it is likely that calculating the $\mathcal{O}(\alpha_S)$ subprocesses $q_1^*\,q_2^* \to g\,V$ and $q_i^*\,g_j^* \to q\,V$ using the $(z,k_t)$-factorisation prescription would account for any significant higher order corrections not already included and so reduce the observed discrepancy without refitting the input integrated PDFs.

\begin{figure}
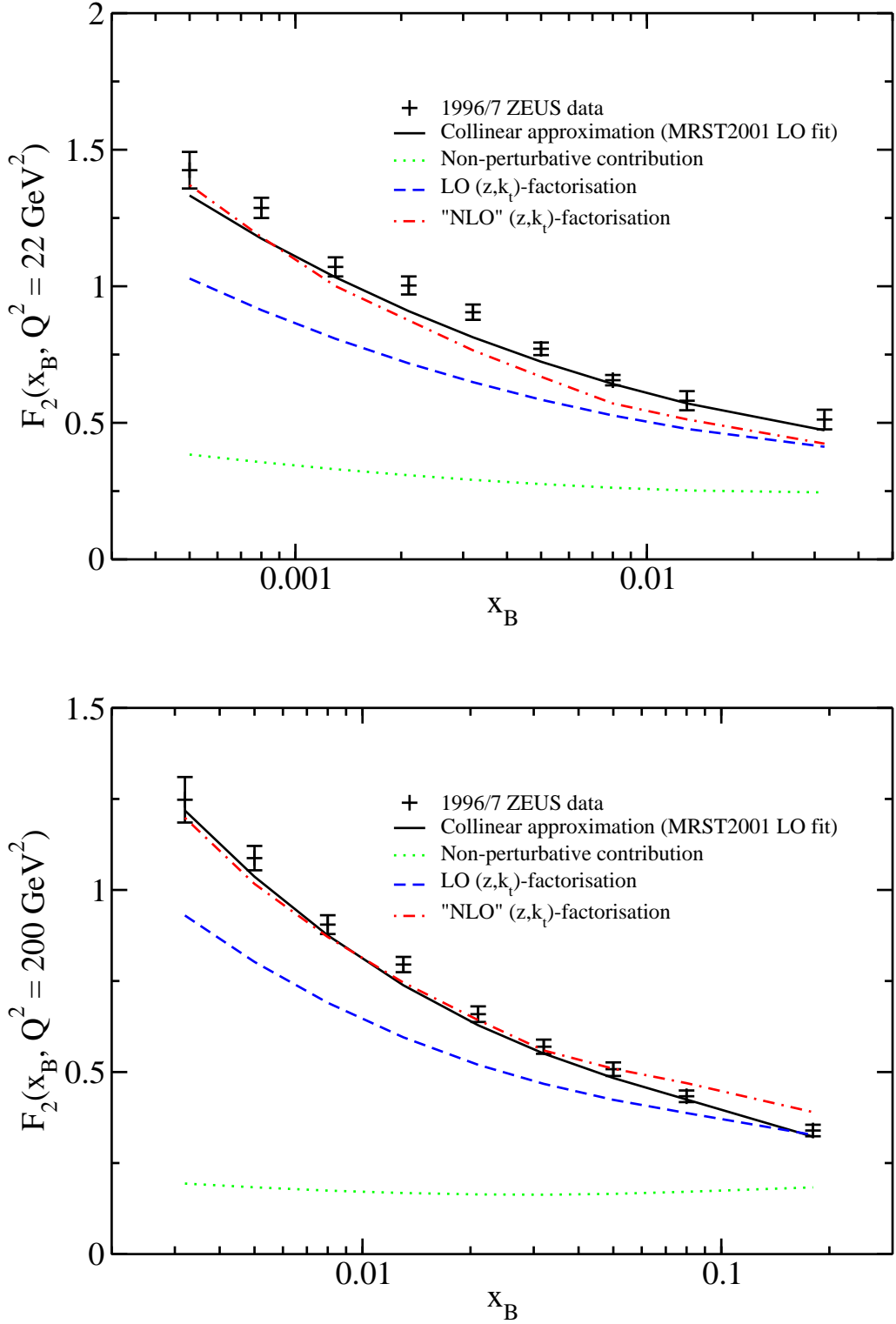

  \centering
  \includegraphics[width=0.8\textwidth,clip]{fig7a.eps}\\[1cm]
  \includegraphics[width=0.8\textwidth,clip]{fig7b.eps}
  \caption{Predictions for $F_2(\xB,Q^2)$ at $Q^2$ values of 22 GeV$^2$ (top) and 200 GeV$^2$ (bottom) in the collinear approximation, where the 1996/97 ZEUS data \cite{Chekanov:2001qu} has been included in the MRST2001 LO fit \cite{Martin:2002dr}, and in the $(z,k_t)$-factorisation approach using the same PDFs as input.  The discrepancy between the data and the LO $(z,k_t)$-factorisation prediction can be eliminated by either refitting the input integrated PDFs or by adding some ``NLO'' contribution.}
  \label{fig:F2}
\end{figure}

These reasons for the small discrepancies discussed in the previous two paragraphs can be illustrated by considering the proton structure function $F_2(\xB,Q^2)$, which is plotted for two $Q^2$ values in Fig.~\ref{fig:F2}.  In the collinear approximation, the LO prediction for this observable comes from $\gamma^* q\to q$
\begin{equation} \label{eq:qpm}
  F_2(\xB,Q^2) = \sum_q e_q^2\,\xB q(\xB,Q^2),
\end{equation}
indicated by the solid lines in Fig.~\ref{fig:F2}, which gives a good description of the data points since this data set was included in the MRST2001 LO global fit.  The LO $(z,k_t)$-factorisation prediction comes from $\gamma^* q^*\to q$ and may be obtained from (85) of \cite{Watt}:
\begin{equation} \label{eq:F2LOzkt}
  F_2(\xB,Q^2) = \sum_q e_q^2\,\xB q(\xB,\mu_0^2)\,T_q(\mu_0^2,Q^2) + \int_x^1\!\dif{z}\,\int_{\mu_0^2}^{\infty}\!\diff{k_t^2}\;\frac{\xB/x}
{1-\xB\beta/x}\;\sum_q e_q^2 f_q(x,z,k_t^2,\mu^2),
\end{equation}
where the Sudakov variables $x=x_+$ and $\beta$ are given in (54) of \cite{Watt} and the factorisation scale $\mu$ is given in (25) of \cite{Watt}.  The predictions of this formula are shown as the dashed lines in Fig.~\ref{fig:F2}, while the first term of \eqref{eq:F2LOzkt}, representing the non-perturbative contributions from $k_t <\mu_0$, is also shown separately as the dotted lines.  There is a clear difference between the predictions of \eqref{eq:qpm} and \eqref{eq:F2LOzkt}, which increases as $\xB$ decreases, due to the extra kinematic factor in the second term of \eqref{eq:F2LOzkt}.  This difference would be eliminated by fitting the input integrated PDFs using \eqref{eq:F2LOzkt}.  Alternatively, a ``NLO'' prediction for $F_2$ may be calculated from the subprocesses $\gamma^*g^*\to q\bar{q}$ and $\gamma^*q^*\to qg$, and can be obtained from (94) of \cite{Watt}.  Here, a lower limit of $\mu_0$ is taken in the $k_t^\prime$ integration, and the $k_t^\prime < \mu_0$ contribution is instead accounted for using the first term of \eqref{eq:F2LOzkt}.  It is seen that these ``NLO'' predictions, shown as the dot-dashed lines in Fig.~\ref{fig:F2}, give almost the same results as \eqref{eq:qpm}.

\subsection{Standard Model Higgs boson production at the LHC}
\begin{figure}
  \centering
  \includegraphics[width=\textwidth,clip]{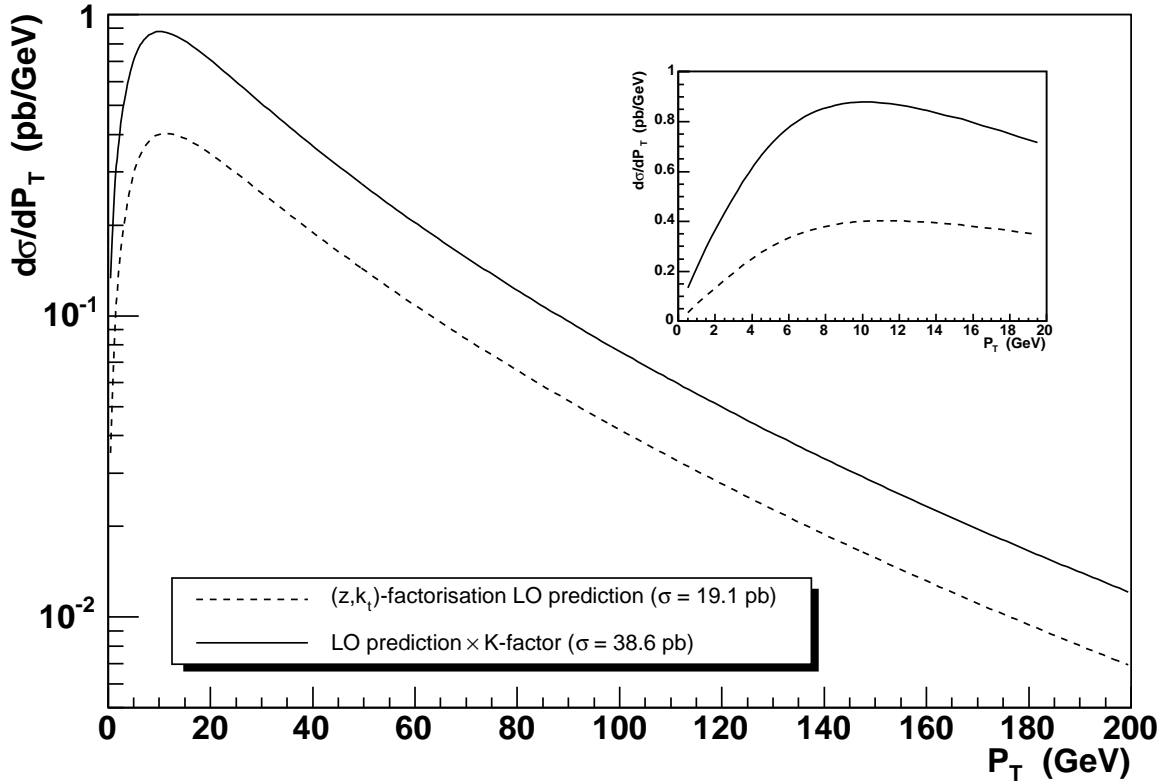}
  \caption{$P_T$ distribution of SM Higgs bosons produced at the LHC with mass 125 GeV.}
  \label{fig:hprod}
\end{figure}

The $P_T$ distribution \eqref{eq:HPT} for SM Higgs bosons of mass 125 GeV produced at the LHC ($\sqrt{s} = 14$ TeV) is shown in Fig.~\ref{fig:hprod}.  To allow direct comparison with the results of \cite{Balazs:2004rd}, we do \emph{not} account for top quark mass effects.  Note that the peak in the Higgs $P_T$ distribution is broader and occurs at a higher $P_T$ than for vector boson production.  This is primarily due to the enhanced $g\to gg$ colour factor ($C_A=3$) compared to the $q\to qg$ colour factor ($C_F=4/3$), resulting in a larger Sudakov suppression at low $P_T$.  By the same reason the $K$-factor \eqref{eq:KggH} is larger.  For $P_T\lesssim M_H$, the $P_T$ distribution is in good agreement with recent, more sophisticated, resummation predictions (see, for example, \cite{Balazs:2004rd}), bearing in mind the spread in the various predictions available due to the different approaches and PDFs used.  However, the peak occurs at a $P_T$ about 1--2 GeV lower than the majority of the resummation predictions (cf.~Fig.~1 of \cite{Balazs:2004rd}).  Evaluating the total cross section by integrating over all $P_T$ gives 38.6 pb, close to the next-to-next-to-leading order (NNLO) calculation which gives 39.4 pb \cite{Balazs:2004rd}.

Note that matrix-element corrections are necessary in parton shower simulations at large $P_T$.  Without such corrections, the \textsc{herwig} parton shower prediction falls off dramatically at large $P_T\gtrsim M_H$ \cite{Balazs:2004rd,Corcella:2004}.  The fact that our prediction is much closer to the fixed-order results than parton shower predictions suggests that we have successfully accounted for a large part of the sub-leading terms.

\section{Conclusions}

We have extended the method of $(z,k_t)$-factorisation using doubly-unintegrated parton distributions \cite{Watt} to hadron-hadron collisions.  The key idea is that the incoming partons to the subprocess have finite transverse momenta, which can be observed in the particles produced in the final state.  This transverse momentum is generated perturbatively in the last evolution step, with a suitable extrapolation for the non-perturbative contribution.  Virtual terms in the DGLAP equation are resummed into Sudakov form factors and angular-ordering constraints are applied which regulate soft gluon emission.  By accounting for the precise kinematics in the subprocess, together with these Sudakov form factors and angular-ordering constraints, we are able to include the main part of conventional higher order calculations.

We used this framework to calculate the $P_T$ distributions of $W$ and $Z$ bosons produced at the Tevatron Run 1.  The predictions gave a very good description of CDF and D{\O} data over the whole $P_T$ range, after multiplying by an overall factor of 1.1--1.2, corresponding to multiplying each DUPDF by a factor $\lesssim 1.1$.  We explained the origin of the need for this extra factor, which should not be regarded as a deficiency of our approach, but rather reflects the fact that the input integrated PDFs should themselves be determined from data using $(z,k_t)$-factorisation.

We also used the framework to calculate the $P_T$ distribution for SM Higgs bosons of mass 125 GeV produced at the LHC.   For $P_T\lesssim M_H$, our simple prescription was found to reproduce, to a fair degree, the predictions of more elaborate theoretical studies \cite{Balazs:2004rd}.

\section*{Acknowledgements}

ADM thanks the Leverhulme Trust for an Emeritus Fellowship.  This work was supported by the UK Particle Physics and Astronomy Research Council and by the Federal Program of the Russian Ministry of Industry, Science and Technology (grant SS-1124.2003.2).

\end{document}